\documentclass{aa}
\usepackage{graphics,amsfonts}

\newcommand{\kms}{km\,s$^{-1}$}
\newcommand{\ve}{$v_{\rm e}\sin i$}
\newcommand{\dsl}{\displaystyle}

\newcommand{\vb}{\mbox{\boldmath${\it V}$}}
\newcommand{\ov}[1]{\overline{V}_{#1}}
\newcommand{\dfrac}[2]{\frac{\dsl#1}{\dsl#2}}
\newcommand{\mo}{\mbox{$\langle V \rangle$}} 
\newcommand{\mt}{\mbox{$\langle V^2 \rangle$}} 
\newcommand{\mh}{\mbox{$\langle V^3 \rangle$}} 

\newcommand{\figps}[1]{\resizebox{\hsize}{!}{\rotatebox{0}{\includegraphics{#1}}}}
\newcommand{\beq}{\begin{equation}}
\newcommand{\eeq}{\end{equation}}

\begin{document}

\title{Variation of the line profile moments for stars pulsating in
distorted oblique non-radial modes}

\titlerunning{Variation of the line profile moments for oblique non-radial pulsators}

\author{O. Kochukhov}

\institute{Department of Astronomy and Space Physics, Uppsala University, SE-751 20, Uppsala, Sweden\\
           \email{oleg@astro.uu.se}}

\date{Received / Accepted}

\abstract{We derive expressions and develop a numerical technique for the analysis of 
the line profile moment variations for stars pulsating in oblique non-radial modes. This represents an extension
of the widely used spectroscopic moment mode identification method to the oblique distorted pulsations observed in 
rapidly oscillating Ap stars. We demonstrate that a non-axisymmetric superposition of the pulsation
and rotation velocity fields results in a qualitatively new behaviour of some of the line profile characteristics.
It is found that for the majority of roAp stars the second moment varies with the
pulsation frequency rather than with its first harmonic even for axisymmetric modes. We also identify 
pulsation observables which do not change during pulsation cycle but are modulated by the stellar rotation
and can contribute to the variability of the stellar spectra averaged over many pulsation cycles.
As an illustration of the new version of the
moment technique, we compute rotational modulation of the pulsational changes of the line profile moments for the
oblique axisymmetric dipolar pulsation modes with different parameters. It is also shown that 
a distortion of the oblique dipolar modes predicted by the recent theoretical studies of the stellar magneto-acoustic
oscillations can be readily diagnosed through the moment analysis. In particular, the shape of the pulsation phase 
modulation for the radial velocity and the second moment is very sensitive to non-axisymmetric pulsation components, whereas the
rotational modulation of the second moment amplitude is best suited to reveal axisymmetric magnetically 
induced distortion of pulsations.
}
\maketitle

\section{Introduction}
\label{intro}

Analysis of the stellar pulsations is a valuable astrophysical technique offering a unique possibility to
constrain the internal structure and evolution of stars. A detailed information about the surface geometry of the
stellar pulsations is a necessary prerequisite for a successful asteroseismic investigation into the physical
properties of the stellar layers. Pulsational perturbations in the surface layers of stars are described by
the spherical harmonic eigenfunctions. The angular degree $\ell$ and the azimuthal number $m$ is
sufficient to fully characterize the surface geometry of non-radial pulsations in normal slowly rotating
stars. 

A range of the mode identification techniques was developed to extract information about the pulsation geometry from
the photometric (Watson \cite{W88}) and, most recently, spectroscopic (Aerts \& Eyers \cite{AE00};
Telting \cite{T03}) observations of the stellar pulsations. The
moment technique, introduced by Balona (\cite{B86}) and further developed by Aerts et al. (\cite{APW92})
and Briquet \& Aerts (\cite{BA03}), is one
of the most widely applied and efficient spectroscopic mode identification methods. It attempts to diagnose
the structure of pulsational perturbations by interpretation of the time variation of the first three 
velocity moments that are linked to the line centroid (radial velocity), line width and skewness (line asymmetry).

In contrast to normal non-radially pulsating stars,
a considerably more complex and challenging problem is posed by the pulsations in rapidly oscillating Ap
(roAp) stars (Kurtz \& Martinez \cite{KM00}). Since their discovery (Kurtz \cite{K82}) roAp stars have drawn 
great attention as the only example of the stellar oscillations in the presence of a global strong magnetic field.
The stellar magnetism is closely linked to the excitation of the high-overtone short-period {\it p-}modes in roAp
stars and, therefore, asteroseismic analysis may contribute to solving the long-standing problem of the origin
and structure of global stellar magnetic fields. 

The pulsational amplitude and phase of the oscillating 
magnetic stars were found to vary synchronously with the rotational modulation of the line of sight magnetic field. A
straightforward phenomenological explanation of this behaviour was given by the \textit{oblique pulsator 
model} (Kurtz
\cite{K82}), which postulated that pulsations are aligned with the oblique magnetic field and modulation of the
pulsational characteristics is a result of the periodic changes in the aspect angle at which pulsations are
observed. In the standard oblique pulsator framework the modes were represented by the axisymmetric dipolar
($\ell=1$, $m=0$) eigenfunctions.  However, detailed photometric monitoring of some of roAp stars (e.g. Kurtz et al. \cite{KWR97}) and
theoretical studies (Shibahashi \& Takata \cite{ST93}; Bigot \& Dziembowski \cite{BD02}; Saio \& Gautschy
\cite{SG04}) have suggested that pulsation modes in roAp stars are significantly distorted by rotation and magnetic field and
cannot be described by a single spherical harmonic function. Recent time-resolved high-resolution
spectroscopic studies of roAp stars (e.g. Kanaan \& Hatzes \cite{KH98}; Kochukhov \& Ryabchikova \cite{KR01a})
also revealed a startling diversity of the pulsational behaviour of individual spectral lines, which was
interpreted to be an interplay between a short vertical length-scale of pulsation waves and the presence of
significant vertical gradients of chemical abundance in the line forming atmospheric regions (Kochukhov \& Ryabchikova 
\cite{KR01a}; Ryabchikova et al. \cite{RPK02}).

These discoveries emphasized a necessity to develop quantitative tools to analyse in detail the surface
geometry of pulsations in roAp stars in order to distinguish the horizontal and vertical pulsation effects. The
measurements and modelling of the line profile moments appear especially well-suited for Ap stars which are
typically very slow rotators. 
However, the latest version of the standard moment method implemented by Briquet \& Aerts (\cite{BA03}) cannot be 
directly applied to this kind of stars because this technique does not account for the mode obliquity and neglects magnetic
distortion of pulsation modes.

The main aim of the present investigation is to develop a numerical technique
suitable for modelling
the moment variations of stars pulsating in distorted oblique non-radial modes. Moreover, we characterize
the main
features of the pulsational variability of moments in the context of modern theories of the magneto-acoustic
stellar oscillations. The rest of the paper is organized as follows. Sect.~\ref{moments} discusses
mathematical formalism and numerical implementation of the extended moment technique. Calculations of the
moment variations are presented in
Sect.~\ref{illustr} for a variety of oblique pulsation geometries and Sect.~\ref{discuss} concludes the paper
with the summary and discussion.

\section{Influence of pulsation and rotation on the line profile moments}
\label{moments}

In the present study we use representation of the pulsation velocity suggested by Kochukhov (\cite{K04a}).
This paper demonstrated that an arbitrary pulsation velocity field corresponding to a mono-periodic linear
stellar oscillation with the angular frequency $\omega$ can be written as
\beq
\vb(\theta,\phi,t) = \vb^{\rm c}(\theta,\phi) \cos{(\omega t)}+ 
                   \vb^{\rm s}(\theta,\phi) \sin{(\omega t)},
\eeq
where $\theta$ and $\phi$ are the spherical coordinates on the stellar surface, whereas $\vb^{\rm
c}(\theta,\phi)$ and $\vb^{\rm s}(\theta,\phi)$ are the vectors describing the surface distribution of the pulsation amplitudes for the
horizontal and vertical velocity fluctuations:
\beq
\vb^c(\theta,\phi) = \left \{ V^{\rm c}_r, V^{\rm c}_\theta, V^{\rm c}_\phi \right \} {\rm~and~}
\vb^s(\theta,\phi) = \left \{ V^{\rm s}_r, V^{\rm s}_\theta, V^{\rm s}_\phi \right \}.
\eeq
The complete expressions of the $V^{\rm c,s}_{r,\theta,\phi}$ velocity components
in terms of the spherical harmonics were given by Kochukhov (\cite{K04a}).

In the general case of a rotating star with oblique non-radial pulsations each oscillation mode is represented by a
superposition of many spherical harmonic components (Bigot \& Dziembowski \cite{BD02}; Saio \& Gautschy
\cite{SG04}) and there are no coordinate transformations which are able to reduce this description to a
single spherical harmonic eigenfunction assumed in the classical moment technique of Balona (\cite{B86})
and Aerts et al. (\cite{APW92}). As a consequence, the six pulsation velocity components 
$V^{\rm c,s}_{r,\theta,\phi}$ have to be defined in any coordinate system through the sums over the spherical harmonic functions 
$Y_{\ell m}(\theta,\phi)$ and their derivatives. Nevertheless,
there still exists a special reference frame, e.g. the one connected with the global magnetic field of
the star, where the pulsational displacement exhibits certain symmetries and the respective mathematical description of 
the velocity field is the least complicated (Saio \& Gautschy \cite{SG04}). We choose this coordinate system, inclined by the angle $\beta$
relative to the stellar rotation axis, to define the local angular coordinates $\theta$ and $\phi$ and to specify 
pulsation amplitude maps $\vb^{\rm c}(\theta,\phi)$ and $\vb^{\rm s}(\theta,\phi)$
in terms of the sums over the spherical harmonic components.

For the purpose of numerical computation of the disk-averaged characteristics of the line of sight velocity the
visible stellar hemisphere is divided into a number of surface zones with approximately equal areas (as in
Piskunov \& Kochukhov \cite{PK02}). Position of each surface element in the 
stellar reference frame is characterized by the latitude 
$\rho$ ($-\pi/2\le\rho\le\pi/2$) and the longitude $l$ ($0\le l\le2\pi$). 

The transformation of an arbitrary vector field from the spherical coordinate system not aligned with the
stellar rotation axis to the Descartes reference frame of the observer was treated by Piskunov \& 
Kochukhov (\cite{PK02}) in the context of modelling the stellar magnetic field. Similar expressions are
applicable for the pulsation velocity field.  The transformation of the vectors $\vb^{\rm c}$ and
$\vb^{\rm s}$ can be written as a series of matrix multiplications employing the operator
$\mathsf{T}(\gamma,\delta)$ defined as
\beq
\mathsf{T}(\gamma,\delta)\equiv
\left(
\begin{array}{ccc}
\cos{\delta} & -\sin{\delta} & 0 \\
\sin{\delta} & \phantom{-}\cos{\delta} & 0 \\
0 & 0 & 1 \\
\end{array}
\right) \times
\left(
\begin{array}{ccc}
1 & 0 & 0 \\
0 & \cos{\gamma} & -\sin{\gamma} \\
0 & \sin{\gamma} & \phantom{-}\cos{\gamma} \\
\end{array}
\right),
\eeq
where the angle $\gamma$ describes a tilt of the polar ($z$) axis of a given coordinate system and the
angle $\delta$ corresponds to a rotation around this axis. Thus, vector components of the pulsation
velocity amplitudes in the observer's coordinate system whose $z$-axis is directed along the line of
sight are evaluated as
\beq
\left(
\begin{array}{r}
V^{\rm c,s}_x(\rho,l) \\
V^{\rm c,s}_y(\rho,l) \\
-V^{\rm c,s}_z(\rho,l) \\
\end{array}
\right) =
\mathsf{T}(-i,0) \mathsf{T}(\beta,\varphi)
\mathsf{T}(\theta-\dfrac{\pi}{2},\phi) 
\left(
\begin{array}{r}
V^{\rm c,s}_\phi(\theta,\phi) \\
-V^{\rm c,s}_r(\theta,\phi) \\
-V^{\rm c,s}_\theta(\theta,\phi) \\
\end{array}
\right).
\label{crd}
\eeq
Here $\varphi\equiv\Omega t$ is the phase angle
corresponding to uniform stellar rotation with the angular frequency $\Omega$ 
and $i$ is the angle between the stellar rotation axis and the line of sight.

The total time-dependent velocity with respect to the observer 
includes the contributions from pulsation 
and rotation:
\beq
V_z(\rho,l) = V^{\rm c}_z(\rho,l) \cos{(\omega t)} + 
              V^{\rm s}_z(\rho,l) \sin{(\omega t)} + V^{\rm rot}_z(\rho,l).
\eeq
The $z$ component of the
rotation velocity is given by
\beq
V^{\rm rot}_z(\rho,l) = \cos{\rho}\sin{(l+\varphi)}v_{\rm e}\sin{i},
\eeq
where \ve\ is the projected rotational velocity.

We use the usual assumption that the intrinsic line profile can be approximated by a Gaussian with variance
$\sigma^2$. This approximation will fail only for a (relatively uncommon) situation of a magnetically 
sensitive line in the spectrum of a strongly magnetic 
and very slowly rotating roAp star. We follow the moment definitions of Aerts et al. (\cite{APW92}) and express
variability of moments in terms of changes
with pulsation frequency $\omega$, as well as its first and second harmonics, $2\omega$ and $3\omega$.
The expressions for the radial velocity \mo, the second moment \mt\ and
the third moment \mh\ thus become
\beq
\langle V \rangle = \langle V_z(\rho,l) \rangle = \ov{100}\cos{(\omega t)}+\ov{010}\sin{(\omega t)},
\label{mnt1}
\eeq
{\renewcommand{\arraystretch}{1.2}%
\beq
\begin{array}{rcl}
\langle V^2 \rangle & = & \sigma^2 + \langle V^2_z(\rho,l) \rangle = \sigma^2 + \frac{1}{2}\left(\ov{200}+\ov{020}+2\ov{002}\right) \\
		    & + & 2\ov{101}\cos{(\omega t)} + 2\ov{011}\sin{(\omega t)} \\
		    & + & \frac{1}{2}\left(\ov{200}-\ov{020}\right)\cos{(2\omega t)}+\ov{110}\sin{(2\omega t)}
\end{array}
\label{mnt2}
\eeq%
}
and
{\renewcommand{\arraystretch}{1.3}%
\beq
\begin{array}{rcl}
\langle V^3 \rangle & = & \langle V^3_z(\rho,l) \rangle + 3\sigma^2\langle V_z(\rho,l) \rangle = 
                          \frac{3}{2}\left(\ov{201}+\ov{021}\right)\\
                    & + & \frac{3}{4}\left(\ov{300}+4\ov{102}+\ov{120}+4\sigma^2\ov{100}\right)\cos{(\omega t)}\\
		    & + & \frac{3}{4}\left(\ov{030}+4\ov{012}+\ov{210}+4\sigma^2\ov{010}\right)\sin{(\omega t)}\\
		    & + & \frac{3}{2}\left(\ov{201}-\ov{021}\right)\cos{(2\omega t)}+3\ov{111}\sin{(2\omega t)}\\
		    & + & \frac{1}{4}\left(\ov{300}-3\ov{120}\right)\cos{(3\omega t)}\\
		    & + & \frac{1}{4}\left(3\ov{210}-\ov{030}\right)\sin{(3\omega t)}.
\end{array}
\label{mnt3}
\eeq%
}

The quantities $\ov{knm}$ represent the surface integrals of various powers of the $V^{\rm c}_z$, $V^{\rm s}_z$ and
$V^{\rm rot}_z$ velocity components and are approximated with the weighted sums over the visible surface
elements
\beq
\ov{knj}(\varphi)\equiv\sum_{i=1}^N W_i \left[V^{\rm c}_z(\rho,l)\right]^k 
                           \left[V^{\rm s}_z(\rho,l)\right]^n
                           \left[V^{\rm rot}_z(\rho,l)\right]^j.
\eeq
Here the weight function $W_i$ is evaluated under the assumption of the linear limb-darkening law
\beq
W_i \equiv \dfrac{3\left(1-u+u\mu_i\right)\mu_i S_i}{\pi(3-u)},
\eeq
where $u$ is the limb-darkening coefficient, whereas $\mu_i$ and $S_i$ are the cosine between the
surface normal and the line of sight and the intrinsic area of the $i$th surface zone respectively. In deriving
expressions (\ref{mnt1}--\ref{mnt3}) we took into account the symmetry of the rotation velocity
field which reduces the $\ov{00j}$ sums to zero for all odd values of $j$.

We emphasize that the moment equations (\ref{mnt1}--\ref{mnt3}) include variability on the two different
timescales: pulsation with the angular frequency $\omega$ and rotation with the angular frequency 
$\Omega$. The latter
is included in the expressions indirectly through the time dependence of the coordinate transformation
defined by Eq.~(\ref{crd}). Taking into account this fundamental background of the problem of investigation
into the nature of oblique non-radial pulsators it is natural to take an advantage of the stellar rotation
and to observe the pulsation velocity field at different aspect angles -- a unique possibility
realized only in roAp stars. Then the mode identification for roAp stars
and the analysis of the distortion of pulsations by
rotation and magnetic field can be carried out by modelling the temporal evolution of the 
pulsation spectra of \mo, \mt\ and \mh\ as a function
of the stellar rotation phase, similar to the studies of the rapid photometric variability of roAp
stars (Kurtz \& Martinez \cite{KM00}). To describe the theoretical rotational modulation of each 
amplitude $V_{\rm x}(\varphi)$ in Eqs.~(\ref{mnt1}--\ref{mnt3})
we compute an equivalent representation in terms of the total time-dependent amplitude $A_{\rm x}(\varphi)$
and the phase $\Phi_{\rm x}(\varphi)$ 
\beq
\begin{array}{rl}
V_{\rm x}(t)&=a_{\rm x}(\varphi) \cos{(n\omega t)}+b_{\rm x}(\varphi) \sin{(n\omega t)} \\
            &=A_{\rm x}(\varphi) \cos{\left[n\omega t+\Phi_{\rm x}(\varphi)\right]}, \\
\end{array}
\eeq
where
\beq
A_{\rm x}(\varphi)=\sqrt{a^2_{\rm x}(\varphi)+b^2_{\rm x}(\varphi)} {\rm~~and~~} 
\Phi_{\rm x}(\varphi)=\arctan{\dfrac{a(\varphi)}{b(\varphi)}}-\dfrac{\pi}{2}.
\eeq

Examination of the moment formulae derived for the oblique pulsators reveals some essential differences
with respect to the moment behaviour in normal pulsating stars. Apart from the expected rotational modulation
of all terms in
the moment amplitudes and phases, the most important feature is the relative amplitudes of the terms contributing
to the pulsational variation of the second moment. In stars where pulsations are aligned with the rotation axis and the pulsation velocity
is of the same order as the rotation one, contribution
of the main frequency term is usually small or comparable to the $2\omega$ term and even reduces to zero for any rotation 
velocity if pulsations are axisymmetric (i.e., in the language of terms of Eq.~(\ref{mnt2}): $\ov{101,011}(\beta=0,m=0)=0$). 
This does not happen for oblique pulsators. For all but the slowest
rotating roAp stars pulsation amplitudes are smaller than the projected rotational velocity. Hence,
$\ov{101,011}(\beta\ne0)\gg(\ov{200}-\ov{020})$
even for an axisymmetric pulsation and we expect to see a strong single
wave variability dominating the pulsation behaviour of \mt\ in most of the roAp stars.

Another interesting feature of the extended moment equations is the rotational modulation of the ``constant'' terms in the expression for
the second moment and the presence of such terms for the third moment. Although changes of these terms on the timescale of the
stellar rotation are relatively small, this effect illustrates nicely a consequence of the non-asymmetric co-addition
of the pulsation and rotation velocity fields and may contribute to the variation of line profiles in the 
regular (averaged over many pulsation cycles) spectroscopic observations of roAp stars.

In comparison
with the classical semi-analytical expressions of the moment variation derived by Aerts et al. (\cite{APW92}),
the numerical scheme introduced here is considerably more general and
straightforward to implement. It is also more flexible due to the possibility to include additional
physical effects (such as stellar surface inhomogeneities) in the weight function $W$. A similar
numerical approach was also followed in the recent modification of the moment method presented by
Briquet \& Aerts (\cite{BA03}), who optimized the moment technique for multiperiodic rotating stars but
did not address the problem of modelling the moment variation for stars with oblique pulsations.

\begin{figure*}[th*]
\figps{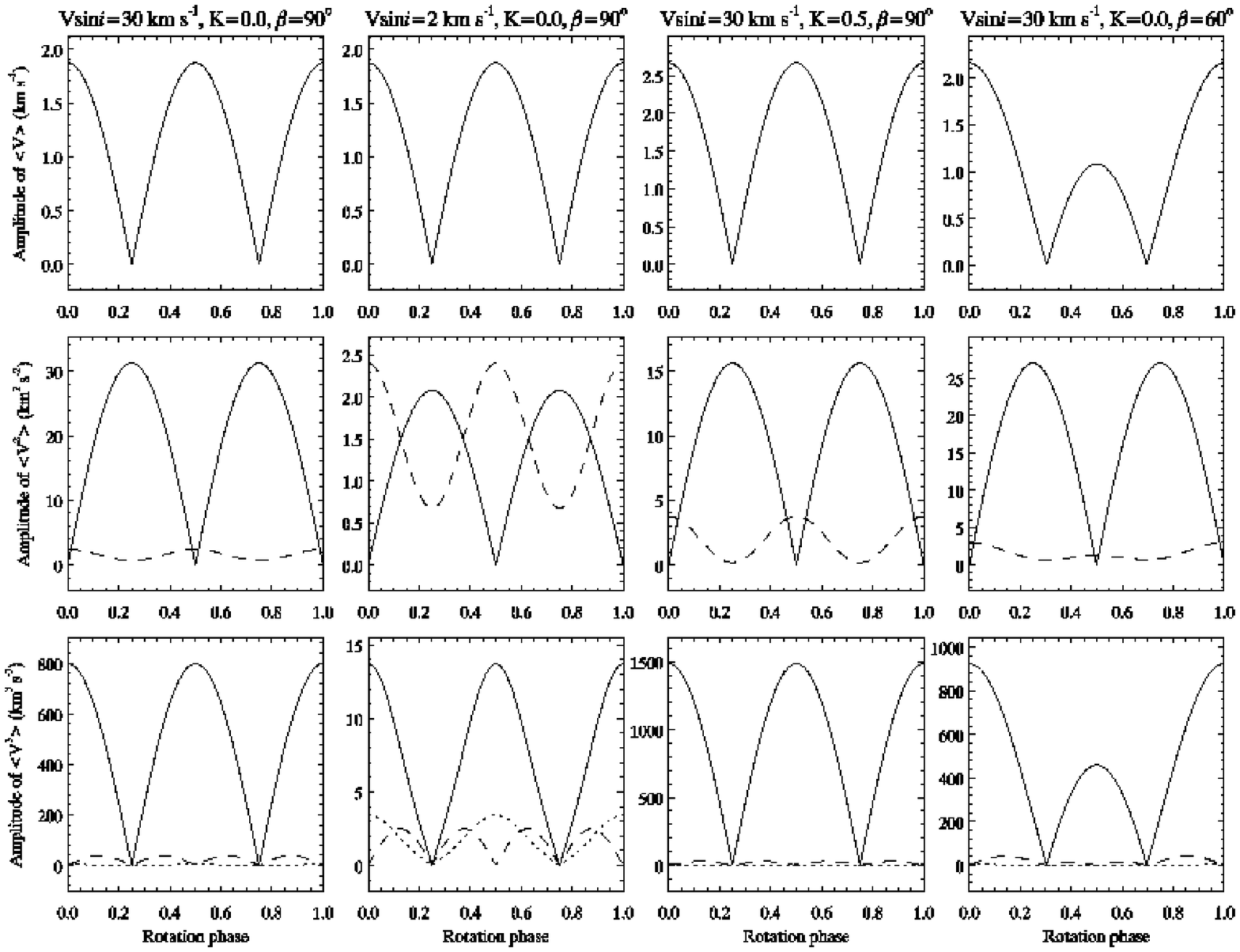}
\caption{Rotational modulation of the pulsation amplitudes of the radial velocity \mo, second moment
\mt\ and third moment \mh\ computed for the oblique dipolar ($\ell=1, m=0$) non-radial mode
with $V_{\rm max}=4$~\kms\ and $i=60\degr$. The other relevant parameters -- the projected rotation velocity \ve, the ratio of the horizontal
to vertical pulsation amplitude $K$ and the mode obliquity $\beta$ -- are indicated at the top of the uppermost panel
in each column of plots. Different curves show amplitudes for the terms changing with pulsation frequency (solid line),
its first (dashed line) and second (dotted line) harmonics.}
\label{fig1}
\end{figure*}

\begin{figure*}[th*]
\figps{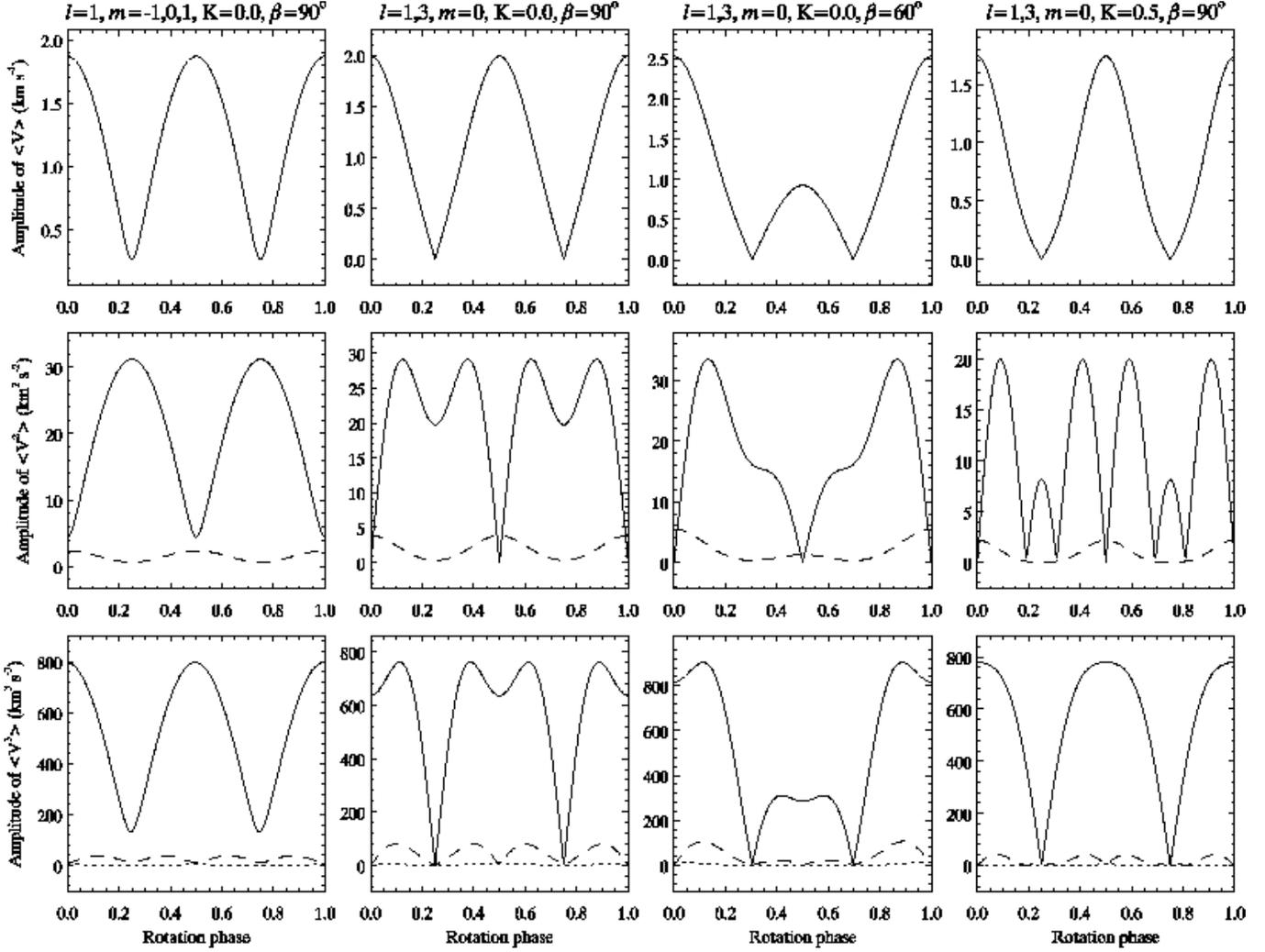}
\caption{Same as Fig.~\ref{fig1} for the amplitude modulation of the moments computed for distorted dipolar modes.
First column: non-axisymmetric dipolar mode with $V_{m=1}=V_{m=-1}=0.1 V_{m=0}$, second to fourth columns: magnetically distorted
axisymmetric mode with $V_{\ell=3}=0.5 V_{\ell=1}$ and different $K$ and $\beta$.}
\label{fig2}
\end{figure*}

\begin{figure*}[th*]
\figps{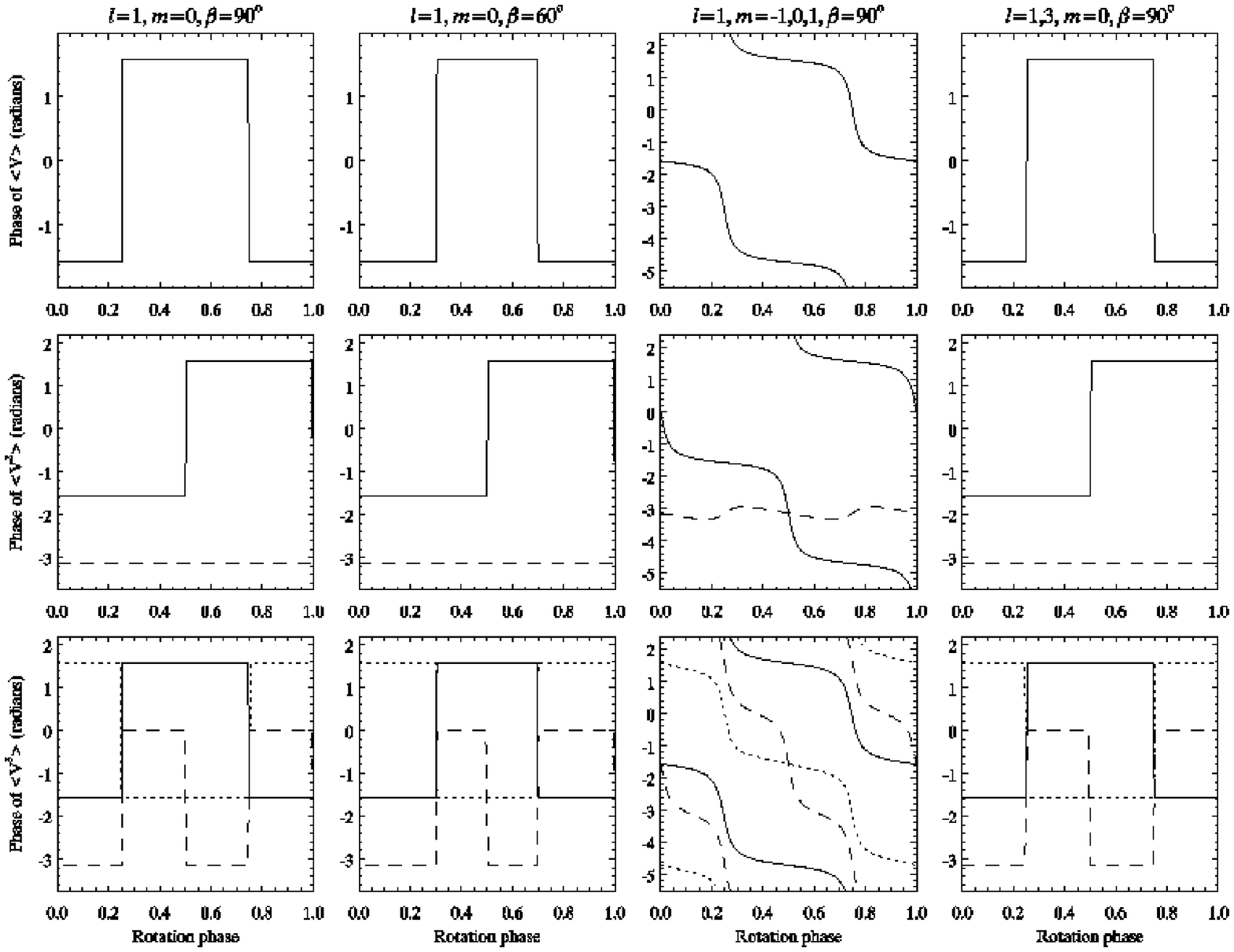}
\caption{Rotational modulation of the pulsation phase of the line profile moments. The format of the figure is similar to that
of Figs.~\ref{fig1} and \ref{fig2}.}
\label{fig3}
\end{figure*}

\section{Illustration of the moment variations}
\label{illustr}

For all calculations of the line profile moments presented in this section we used a 1876-element
surface grid, linear limb-darkening coefficient $u=0.5$ and adopted an intrinsic profile with
$\sigma=5$~\kms. The maximum surface amplitude of the pulsation velocity was set to 
$V_{\rm max}={\rm max}\left(\sqrt{(V^{\rm c})^2+(V^{\rm s})^2}\right)=4$~\kms. These values
correspond to a typical limb-darkening, thermal Doppler line width and pulsational amplitude
of the late-A rapidly oscillating peculiar stars.
The reference values of the other relevant parameters were chosen to be close to those of the
prototype roAp star HR\,3831 (Kochukhov et al. \cite{KDP04}; Kochukhov \cite{K04b}): \ve\,=30\,\kms,
$i=60\degr$, $\beta=90\degr$. We also attempt to test an effect of a significant horizontal pulsation
component by comparing calculations with the ratio of the horizontal to
vertical pulsation amplitude $K=0.5$ (cf. Saio \& Gautschy \cite{SG04}) with the case of purely vertical pulsations.

Figures \ref{fig1}--\ref{fig3} present rotational modulation of the 
amplitude and phase of each term in the variation of the first three moments.

\subsection{Axisymmetric dipolar modes}

The classical oblique pulsator model is represented by a pure axisymmetric dipolar ($\ell=1$, $m=0$) 
mode inclined with respect to the rotation axis (Kurtz \cite{K82}). If the mode obliquity is close to
$90\degr$, the radial velocity variation reaches maximum at the two phases during rotation cycle
(Fig.~\ref{fig1}), corresponding to the moments of the best visibility of each of the two pulsation
poles. The amplitude modulation is clearly non-sinusoidal, as it features sharp minima and broad maxima.

In agreement with the qualitative discussion in Sect.~\ref{moments}, in the case of a moderately rapid 
rotation (\ve\,$\gg$\,$V_{\rm max}$) variation of \mt\ is dominated by the main frequency term and
occurs with the $\pi/2$ radian shift in rotation phase relative to the modulation of radial velocity. As
demonstrated by Fig.~\ref{fig1}, the variation with twice the pulsation frequency becomes dominant only when the maximum
pulsation velocity exceeds \ve. Modulation of the respective term in the amplitude of the second moment
is sinusoidal and occurs out of phase with respect to modulation of the main frequency term. Harmonic
contributions to the pulsational changes of the third moment are always small.

A significant horizontal pulsation displacement with the ratio of the horizontal to
vertical pulsation amplitude $K=0.5$ has no effect on the shape of the rotational modulation of the
amplitudes of moments. At the same time, horizontal motions have a profound effect on the maximum 
amplitude achieved during rotation cycle: the \mt\ amplitude decreases by roughly a factor of 2,
whereas the maximum amplitude of the third moment increases by the same amount.

For an intermediate ($\beta=60\degr$) obliquity of the axisymmetric dipolar mode we 
preferentially sample one of the pulsation poles. Consequently, one of the maxima of the amplitude of 
\mo\ and \mh\ decreases. The rotational modulation of the amplitude of the second moment
is weakly affected.

The pulsational phase modulation of the axisymmetric oblique dipolar modes is illustrated in the first
and second columns of Fig.~\ref{fig3}. The appearance of the phase modulation curves does not depend 
on \ve\, and the presence of the horizontal pulsation fluctuations. The phase of radial velocity
exhibits two $\pi$ radian phase jumps characteristic of the oblique dipolar pulsation. The rapid change
in phase occurs when pulsation equator is passing through the plane defined by the rotation axis and
the line of sight. In contrast, the $\pi$ radian jump in the phase of the main frequency term of the second moment
occurs when pulsation poles are facing the observer. The phase of the first harmonic term is not modulated by
the stellar rotation, whereas the respective term in the pulsational changes of the third moment shows
four jumps during rotation cycle.

An effect of the moderate mode obliquity on the phase modulation is illustrated in the second column of
Fig.~\ref{fig3}. A decrease in the visibility of one of the pulsation poles does not change the
overall character of the phase modulation, but modifies the times of the $\pi$ radian jumps, so the
latter are no longer equidistant in rotation phase.

\subsection{Non-axisymmetric dipolar modes}

In their recent theoretical study Bigot \& Dziembowski (\cite{BD02}) have introduced a modified oblique
pulsator model of roAp stars, which represented a significant departure from the classical axisymmetric
dipolar pulsation geometry. Bigot \& Dziembowski (\cite{BD02}) suggested that an accurate
non-perturbative treatment of the interaction between stellar rotation, pulsations and magnetic field
may lead to a representation of the pulsation eigenmodes with a complex superposition of the spherical harmonic
functions which contain substantial non-axisymmetric components. Furthermore, strong effects due to the
centrifugal force are expected to break locking of the pulsation axis with the dipolar magnetic field,
so that pulsations are aligned with neither the magnetic or rotation axes. However, this entirely new
geometrical picture of the roAp oscillations remains somewhat speculative, as Bigot \& Dziembowski
(\cite{BD02}) gave no general predictions for the amplitude of the non-axisymmetric contribution to the
pulsation velocity field and failed to consider distortion of the pulsation modes due to a strong 
magnetic field typical for the majority of roAp stars.

It turns out that analysis of the moment variation provides a very sensitive diagnostic of a
non-axisymmetric pulsation. In this section we investigate an effect of small non-axisymmetric ($m=\pm
1$) contributions to the dominant oblique dipolar mode. In the pulsation reference frame the amplitude
of the non-axisymmetric components is set to 10\% of the velocity due to the $\ell=1$, $m=0$ component.
This distortion of the dipolar pulsation has a marginal effect on the modulation of the amplitudes of
moments (Fig.~\ref{fig2}, first column). Rotational changes of the radial velocity amplitude become
smoother and the amplitude does not drop to zero at phases 0.25 and 0.75. At the same time, even very
small non-axisymmetric components have a dramatic effect on the phase modulation curves, as illustrated
in the third column of Fig.~\ref{fig3}. The $\pi$ radian jumps vanish and are replaced by the steep but
continuous phase changes. This phenomenon is a straightforward consequence of the surface distribution
of the pulsation velocity phase. All axisymmetric modes have constant phase with the $\pi$ radian jumps at
the node lines, whereas the non-axisymmetric pulsation components are characterized by a continuous change
of the pulsation phase from one surface zone to the next. This is readily reflected in the disk averaged
velocity moments and allows to identify a very low amplitude non-axisymmetric component of oblique pulsations.

\subsection{Magnetically distorted dipolar modes}

In another important theoretical development Saio \& Gautschy (\cite{SG04}) have studied the effect of
a strong magnetic field on the geometry of oblique non-radial modes. The stellar rotation and the
resulting non-axisymmetric pulsational contributions were neglected, but calculations were performed
for the magnetic field strength of up to 9~kG. In the oblique pulsator model of Saio \& Gautschy (\cite{SG04})
pulsations remain axisymmetric and are aligned with the magnetic field axis. The effect of dipolar
magnetic field is to constrain oscillations to the surface areas close to magnetic poles. Thus, an
initially dipolar pulsation is distorted in such a way that it is described by a sum of the odd
axisymmetric pulsation components. We computed moment variation for such a pulsation model assuming
that the octupolar, $\ell=3$, pulsation has a 50\% smaller amplitude compared to the $\ell=1$ component
and contribution of other harmonics is negligible.  

The amplitude modulation of the line profile moments computed for magnetically distorted oblique
pulsations is presented in Fig.~\ref{fig2} (columns 2--4). The most significant effect of the $\ell=3$
component appears in the modulation of the second moment amplitude. The respective curve becomes more
complex and two additional minima develop at the rotation phases 0.25 and 0.75. Variation of the second
moment is also quite sensitive to the obliquity of the mode and to the presence of the horizontal
pulsation motions (predicted in the theory of Saio \& Gautschy \cite{SG04}). Thus, magnetic distortion
and related enhancement of the horizontal amplitude of pulsations in roAp stars can be most efficiently
studied with the analysis of the rotational modulation of the variation of the second moment of 
spectral line profiles.

Despite important influence on the moment amplitudes, a moderate contribution of the octupolar axisymmetric
spherical harmonic considered here does not change modulation of the phase curves (Fig.~\ref{fig3},
fourth column): it remains identical to the phase modulation obtained for the pure oblique dipolar mode.

\section{Summary and discussion}
\label{discuss}

Recent advances in theoretical studies of the stellar magneto-acoustic pulsations emphasize a necessity
to obtain detailed empirical information about the structure of the pulsational disturbances in the
atmospheres of roAp stars. Until recently, analyses of this interesting class of non-radial pulsators
were limited to the time-resolved photometric observations. Interpretation of the broad-band
photometric variability turned out to be a challenging task because of the difficulty to discriminate
between various effects contributing to the luminosity variation, the anomalous wavelength dependence
of the photometric pulsation amplitudes and phases (Watson \cite{W88}), non-linearity and non-adiabatic
effects (Medupe \cite{M02}). As a result of these complications, little progress has been made in
understanding the vertical and horizontal geometry of the roAp pulsations beyond the generic oblique
pulsator model of Kurtz (\cite{K82}). In fact, the recent study by Saio \& Gautschy (\cite{SG04}) has
demonstrated that, due to the averaging over the visible stellar hemisphere, the photometric observables
are very weakly sensitive to the details of the pulsation geometry and turn out to be practically
useless in testing predictions of elaborate theoretical models.

The newly emerged field of the
observation and interpretation of the pulsational line profile variations in roAp stars (Kochukhov \&
Ryabchikova \cite{KR01a}) offers a more
promising research direction. Due to their simple relation to the pulsational displacement, the
spectroscopic observables provide a more direct, reliable and general approach to modelling the roAp
pulsations compared to interpretation of the photometric pulsational data, which requires some
restrictive assumptions about the physics of pulsations to be made. The most detailed and
straightforward way to reveal the pulsation geometry of roAp stars is to construct a two-dimensional
image of the pulsation velocity field from the line profile variation using the pulsation Doppler imaging
technique (Kochukhov \cite{K04a}, \cite{K04b}). However, this sophisticated modelling requires spectroscopic observational
data of a superb quality and is limited to moderately rotating roAp stars. An alternative possibility is
to apply the moment method (Briquet \& Aerts \cite{BA03}) which is based on interpretation of the low-order
moments of the absorption line profiles and can be applied to very slowly rotating stars.

In this paper we presented an investigation of the diagnostic potential of the moment method in its
application to the oblique distorted stellar pulsations. We extended the moment technique to deal
with the stellar pulsation velocity field with the symmetry axis other than the axis of stellar
rotation and developed a numerical method for the computation of the rotational modulation of amplitudes
and phases of different terms in the variation of the line profile moments. Calculations with our
novel version of the moment technique reveal some important differences in the behaviour of the line profile
moments in normal and oblique pulsators. In particular, for all but the slowest rotating roAp stars
with slightly distorted dipolar modes we expect to observe a single wave variation of \mt\ (the
second moment) during pulsation cycle\footnote{The presence of such single wave second moment variation in
the extremely slowly rotating roAp star $\gamma$\,Equ (Kochukhov \& Ryabchikova \cite{KR01a}) is thus
confirmed to be a signature of a non-axisymmetric non-dipolar pulsation.}. Furthermore, a
non-axisymmetric superposition of the pulsation and rotation velocity distributions results in new
terms in the variation of the higher line profile moments which are modulated by the stellar rotation but
are constant on the pulsation timescale. 

Calculation of the line profile moments for various pulsation
geometries of roAp stars predicted by the theoretical studies demonstrates that analysis of the rotational
modulation of variablity of the first two moments (\mo\ and \mt) is sufficient to
diagnose various non-axisymmetric and axisymmetric distortions of the basic oblique dipolar pulsation
geometry and allows to carry out an extremely sensitive verification of the theories attempting to
describe an interaction between pulsation, magnetic field and stellar rotation. For instance, the non-axisymmetric
modified oblique pulsator theory of Bigot \& Dziembowski (\cite{BD02}) can be very efficiently tested
for a large number of roAp stars by using observations of the rotational modulation of the pulsation phase
of radial velocity and higher line profile moments. The presence of a smooth phase modulation instead
of the $\pi$ radian jumps is a unique signature of non-axisymmetric modes. It reveals a weak
non-axisymmetric contribution even in the background of a much stronger axisymmetric component. 
Similarly, the presence of the odd
axisymmetric harmonic components arising due to the influence of the global dipolar magnetic field (Saio \&
Gautschy \cite{SG04}) can be verified by the observation of the second moment variation.

The promising diagnostic potential of the new moment technique can be realised through interpretation of
the high resolution observations of the line profile variation in oblique pulsators. The bright, high-amplitude
roAp stars with short rotation periods ($\la$10$^{\rm d}$) are the best candidates for the application of the
moment method because their surface pulsation velocity structure can be observed from different
aspect angles during a few night observing run. Observational data for the prototype roAp star HR\,3831 have
already been acquired by Kochukhov \& Ryabchikova (\cite{KR01b}). The time-series analysis of this object and
interpretation of the pulsational spectral variation with the moment technique will be presented in
forthcoming paper.

\begin{acknowledgements}
This work was partially supported by the post-doctoral stipend from the Swedish Research Fund and by the
Lise Meitner fellowship from the Austrian Science Fund (FWF, project No. M757-N02).
\end{acknowledgements}

\end{document}